\documentclass{template}

\usepackage{graphicx}%
\usepackage{multirow}%
\usepackage{amsmath,amssymb,amsfonts}%
\usepackage{amsthm}%
\usepackage{mathrsfs}%
\usepackage[title]{appendix}%
\usepackage{xcolor}%
\usepackage{textcomp}%
\usepackage{manyfoot}%
\usepackage{booktabs}%
\usepackage{algorithm}%
\usepackage{algorithmicx}%
\usepackage{algpseudocode}%
\usepackage{listings}%

\begin{document}

\pagestyle{fancy}
\rhead{}

\title{Hybrid electroluminescence device for on-demand single photon generation at room temperature}

\maketitle

\author{Aleksander Rodek*}
\author{Mateusz Hajdel}
\author{Kacper Oreszczuk}
\author{Anna Kafar}
\author{Muhammed Aktas}
\author{Łucja Marona}
\author{Marek Potemski}
\author{Czeslaw Skierbiszewski}
\author{Piotr Kossacki}

\begin{affiliations}

A. Rodek, K. Oreszczuk, M. Potemski, P. Kossacki\\

Institute of Experimental Physics, Faculty of Physics, University of Warsaw, Pasteura 5, Warsaw, 02-093, Poland\\

Email Address:\\
aleksander.rodek@fuw.edu.pl, kacper.oreszczuk@fuw.edu.pl,\\ marek.potemski@lncmi.cnrs.fr, piotr.kossacki@fuw.edu.pl\\

M. Hajdel, A. Kafar, M. Aktas, Ł. Marona, C. Skierbiszewski\\
Institute of High Pressure Physics, Polish Academy of Sciences, Soko\l{}owska 29/37, Warsaw, 01-142, Poland\\
Email Address:\\
hajdel@unipress.waw.pl, ak@unipress.waw.pl, lucja.marona@unipress.waw.pl, czeslaw@unipress.waw.pl \\
M. Potemski\\
Laboratoire National des Champs Magn\'{e}tiques
Intenses, CNRS-UGA-UPS-INSA-EMFL, 25 Av. des Martyrs Grenoble, 38042, France\\
CEZAMAT-CENTERA2 Labs, Warsaw University of Technology, Koszykowa 75, 00-662 Warsaw, Poland

\end{affiliations}


\begin{abstract}

Recent research focused on single photon emitters (SPEs) hosted by layered semiconductors, particularly hexagonal boron nitride (hBN), has revealed a promising alternative to quantum dots (QDs) for generating single, indistinguishable photons. hBN-based SPEs offer lower material costs, room temperature emission, and easy integration into potential optoelectronic devices due to the layered structure of the host crystal. This work presents compact hybrid electroluminescence devices, in which GaN laser diodes (LDs) are used for bottom-to-top excitation of hBN nanoflakes deposited on the laser facets. Our approach circumvents the issue of electroluminescence generation from hBN and provides access to the SPE's signal without optical driving by an external laser. Using laser diodes upgraded with Bragg reflectors a room-temperature generation of single photons from hBN is confirmed by an 80\%-dip in their g$^{(2)}$ second-order correlation. The on-demand emission of single photons at room temperature is demonstrated by driving the laser diodes in pulsed operation, with confidence supported by a measured g$^{(2)}$(0) value of 0.37.

\end{abstract}

\section{Introduction}\label{sec1}
Solid-state single photon emitters (SPEs) are nowadays one of the most promising platforms for providing commercially viable devices that could be utilized in emerging technologies, which take advantage of the quantum properties of light, e.g. quantum information storage, communications, and sensing \cite{ObrienScience07, AtatüreNatureRevMat18, ShastriNature21}.  Significant progress has been made in the last decade, particularly for the system of quantum dots (QDs) based on III-V materials, which resulted in the fabrication of nanostructures able to deterministically emit indistinguishable single photons with high efficiency\cite{AharonovichNaturePhot16, SenellartNaturenanotech17, GurioliNAtureMat19, LiuNatureNanotech19}. The QD-based single photon sources are already commercially available, underscoring their practicality, but their production is not trivial.  Future progress in implementing quantum technologies on a larger scale requires solutions that are, among others, production-time-effective, reliable, and reproducible. All these issues are particularly obstructive for the platform that ultimately depends on the manual selection of the most optimal stochastically-produced single-photon-emitting sites, or frequently even the entire photonic structures\cite{HennessyNature07, SomaschiNaturePhot16, UppuScience20}.  In consequence, there is currently a strong drive to identify alternatives to the QD-based single photon sources, which has already led to many breakthrough works concerning single-photon emission from layered semiconductor materials: transition metal dichalcogenides (TMDs) and hexagonal boron nitride (hBN)\cite{VasconcellosPSSB22, CianciNanoFut24}. 
 Indeed, various processing methods have been recently developed, allowing for the deterministic manufacture of efficient single-photon-emitting sites in TMDs and hBN\cite{VasconcellosPSSB22, CianciNanoFut24}. Particularly, the latter material offers an important opportunity to observe quantum emitters under ambient conditions\cite{KoperskiOptComm18, KoperskiSciRep21}. Simultaneously, full integration of layered materials into various optoelectronic devices, e.g. light emitting diodes requires efficient electrical driving of their luminescence signal. However, the essential step of controlled doping of TMDs and hBN remains still a largely unresolved challenge\cite{WangAdvMat18}. As such the up-to-date demonstrations of electroluminescence devices based on layered materials almost universally depend on time- and work-extensive manufacturing processes like the design and lithography of electrical contacts or the manual stacking of flakes with micrometer precision\cite{SundaramNanoLett13, ChengNanoLett14, WangNanoLett017,WangAdvMat18, OgawaMaterials23}. In that context, an alternative approach has been recently demonstrated that utilized hybrid devices, where monolayers of TMDs were directly deposited on top of GaN-based blue-emitting LEDs \cite{OreszczukNanoscale22}. This bottom-to-top excitation scheme has proved able to generate not only the standard excitonic spectra of TMD luminescence but also single-photon-emission without the necessity of driving by an external laser, albeit exclusively at cryogenic temperatures. Such electrically-driven single photon emission, which could ideally operate at room temperature, would thus greatly reduce the size and operational complexity of any potential device adding to its scalability on an industrial level. 
Here we present a prototype device consisting of a GaN/InGaN laser diode for the bottom-to-top excitation of commercially available hBN nanocrystals deposited directly on laser facets. We demonstrate the generation of single-photon-emission in ambient conditions both under continuous-wave (CW) excitation and on-demand pulsed driving schemes. Importantly, the operation of our single photon source has been made viable with the amended design of InGaN/GaN laser diodes, which have been equipped with specially developed bandpass filters (Bragg reflectors).

\section{Results}\label{sec2}
\subsection{SPE hybrid device}\label{subsec21}
\begin{figure}[H]
    \centering
    \includegraphics[width=0.5\textwidth]{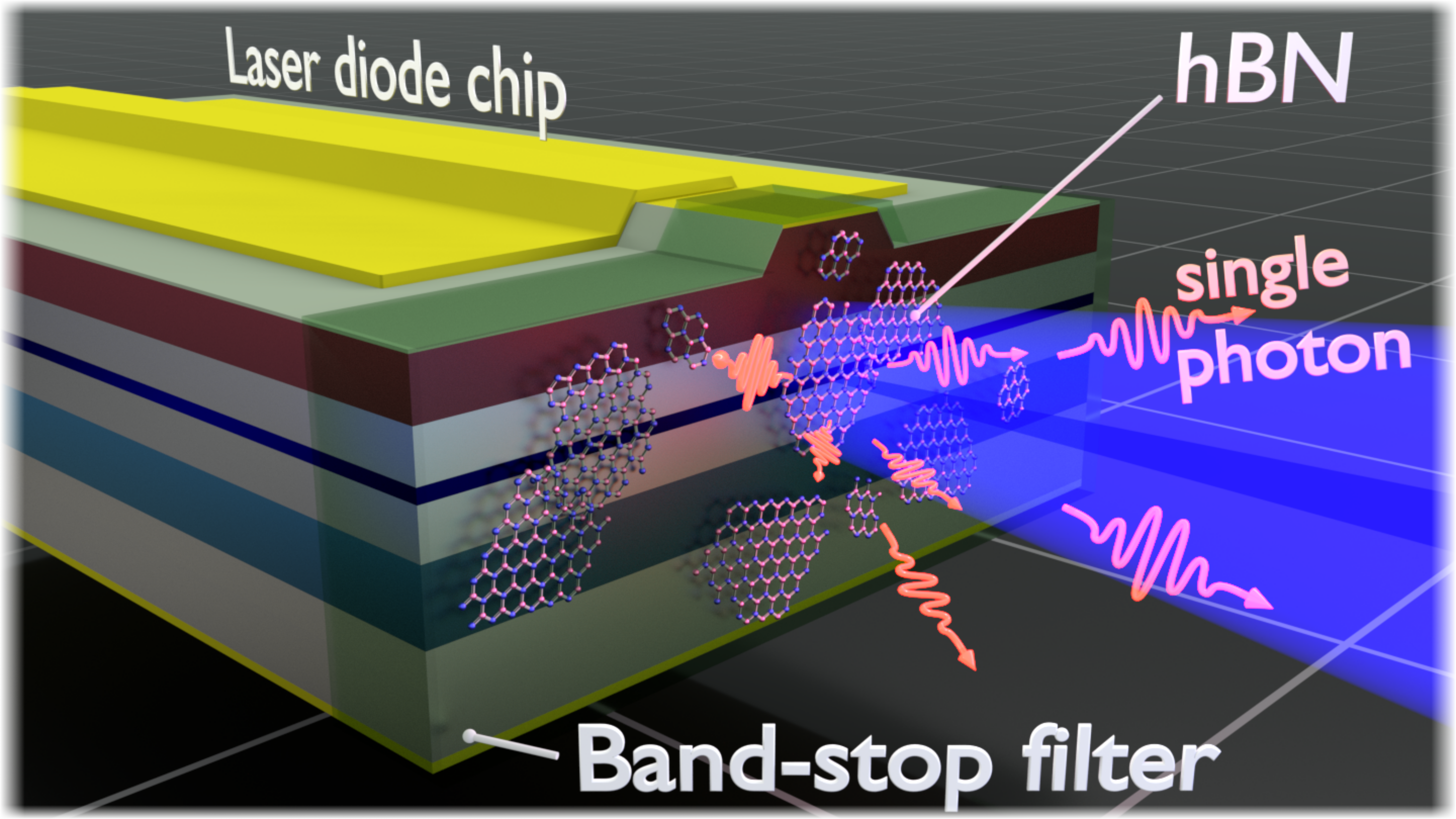}
    \caption{Schematic illustration of the hybrid single photon emitter device}
    \label{fig:0}
\end{figure}

The essential components of our device include: the InGaN/GaN-based laser diode (LD) \cite{stanczyk, Aktas} serving as the excitation source, a band-stop filter, and commercially available nanocrystals of hBN. Such hBN crystal powder has been used in different works in recent years as an easily obtainable source of single photon emitters, which can survive at room temperature \cite{KoperskiOptComm18}.
Narrow emission lines observed in its photoluminescence signal may span the entire visible to near-infrared spectral range\cite{KoperskiOptComm18, Preuß2DMat21, BadrtinovSmall23}. The visualization of the device concept is shown in the Fig.\,\ref{fig:0}

\begin{figure}[H]
    \centering
    \includegraphics[width=0.48\textwidth]{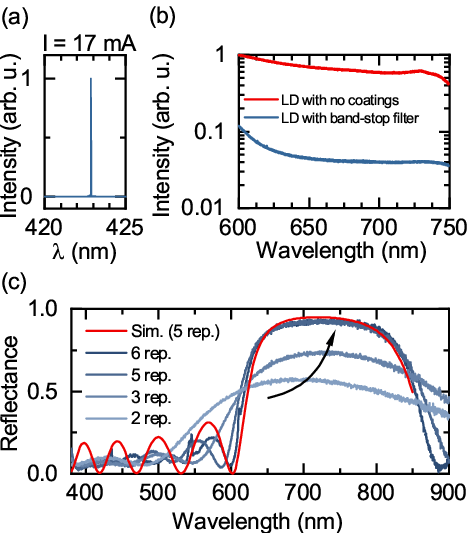}
    \caption{(a) High-resolution lasing spectra of LD near threshold with driving current I=17 mA. (b) Comparison of the long-wavelength spontaneous emission signal of the LD with and without the reflective coating for the same driving conditions. The spectra were collected using FHR1000 Horiba Jobin Yvon 1-meter-long spectrometer in a Gaussian beam telescope setup. All presented measurements were done in continuous wave operation mode with temperature controlled stage at 20\textdegree C (c) Simulated and measured reflectance of the band-stop filter with 2, 3, 5, and 6 repetitions of MgF$_{2}$ and Ta$_{2}$O$_{5}$ dielectric stack.}
    \label{fig:1}
\end{figure}

There are two fundamental requirements concerning the excitation source for it to be suitable for the on-demand optical driving of the SPEs in hBN. Firstly, the optical power density has to be rather large compared to TMDs, e.g. WSe$_2$. Previous reports show that typically 1-2 mW$/\mu m^2$ of optical power have to be delivered to produce the single photon emission at room temperature \cite{KoperskiOptComm18}. Secondly, it should be able to generate excitation impulses with a temporal width lower than the typical lifetime of the single photon emitter. GaN/InGaN-based LD meets all of the requirements given above with additional advantages related to this device family such as high reliability and compact size. The characteristics of the LD and Bragg filters used in the hybrid electroluminescence device are shown in Fig.\,\ref{fig:1}.

In continuous wave operation, LD delivers 100mW of optical power at 115mA of driving current at room temperature (see supplementary data for details). Given the size of the lasing area below $5 \mu m^2$ the device far exceeds the intensity required for exciting the SPEs in hBN ($\sim$ 1-2 mW per $\mu$m$^2$). In particular, this high margin allows us to reduce the heating effects, for example by lowering the laser duty cycle in a pulsed operation mode. For such a case, the necessary average power of $\sim$2mW would correspond to a duty cycle of single percents. The laser wavelength is 422.8nm, which is suitable to efficiently excite hBN SPEs.
Except for the laser emission signal presented in Fig.\,\ref{fig:1}(a), the nitride LD also emits photons in a spontaneous electroluminescence process, with their wavelengths spanning from the visible spectrum up to the near IR. Above the lasing threshold this spontaneous broad-band emission is orders of magnitude weaker than the intensity of the laser radiation. It is, however, strong enough to influence the correlation measurements of single photons. Therefore it is crucial to provide efficient spectral filtering of this addition. For this reason, we developed a laser facet mirror coating that acts as a band-stop filter between 600 and 800 nm. This wavelength range corresponds to the long-wavelength limit of the SPEs obtainable from hBN. It has been chosen to minimize the noise-to-signal ratio related to either the spontaneous emission of the diode, which is additionally filtered by the mirror coating, or the broad emission of the hBN \cite{KoperskiOptComm18}. Both of these contributions display pronounced extinction with the increasing wavelength. 

The band-stop filter mirror was deposited on top of the front facet of the LD. It consisted of a stack (5 or 6 repetitions) of a low refractive index MgF$_2$ (with 129.4 nm thickness) and a high refractive index Ta$_2$O$_5$ (thickness 91.6 nm) layers. In Fig.\,\ref{fig:1}(c) the simulation of the reflectance spectrum of the mirror stack repeated 5 times is presented together with the experimental data for 2, 3, 5, and 6 repetitions. The data displays a saturating behavior of the maximal obtainable value of reflection with the increasing number of layers. A total of 5 repetitions of mirror coatings were used in the final device and showed sufficiently high reflection in the red spectrum regime (up to 92\%) and reasonably beneficial low reflection for the lasing wavelength (5\%). In Fig.\,\ref{fig:1}(b) we present a comparison of the long wavelength spectra of two LDs originating from the same epitaxial process with and without the band-stop filter. For the coated LD the signal at 740 nm drops to 7.2\% of that of the LD with no coatings.

Importantly, we also observe a near 90$\%$ reflectance value in the spectral range of 650-700 nm even up to a relatively high collection angle of 40 degrees in both polarizations(\textit{see supplementary}). The back side of the LD facet was additionally coated with a standard highly reflective dielectric mirror. 

After the deposition of the filters, the coated LDs were mounted in standard TO-56 laser mounts ready for the process of hBN placement.

\subsection{Room-temperature single-photon emission}\label{subsec22}
The straightforward method of obtaining well-isolated single-photon emitting spots is to distribute hBN powder on a polydimethylsiloxane (PDMS) film, which is then stamped onto a substrate surface, very often on Si/SiO$_2$ wafers, and on laser facets in our case. After removing the PDMS film, a certain amount of the hBN powder grains remain on the substrate surface, with their density depending on the initial distribution of hBN on the PDMS film. This process can be further tuned by additional stamping and is sufficient to isolate the SPE sites effectively. Such a simple approach has been applied in the fabrication of our device, while more elaborate techniques, such as large-scale deterministic positioning of hBN nanocrystals\cite{Preuß2DMat21} by capillary assembly, may be advantageous \cite{BoguckiLSA20, PreußNanoLett23} to explore in the future.

\begin{figure}[H]
    \centering
    \includegraphics{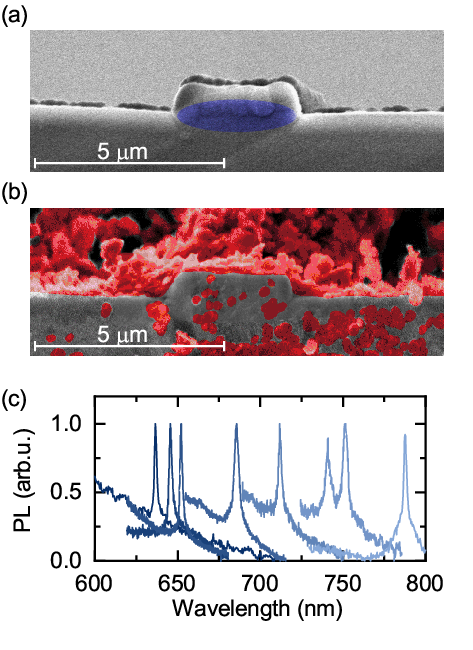}
    \caption{SEM (Scanning Electron Microscope) images of the diode laser facet before (a) and after (b) the deposition of hBN nanopowder. Blue: lasing area of the laser diode. Red: deposited hbn nanopowder. (c) Normalized photoluminescence of hBN narrow emission lines measured at room temperature excited by the laser diode device. Each line spanning the spectral range of 600-800 nm was obtained after a different deposition. Excitation power was in the range of 100-700 $\mu$W. }
    \label{fig:3}
\end{figure}

In Fig.\,\ref{fig:3} we show SEM images of the laser facet before (a) and after (b) the hBN powder deposition. The lasing area is marked in blue, while the hBN nanocrystals are marked in red, displaying their presence on top of the excitation region. In the CW photoluminescence measurements, the laser diode was powered by rectangular electric pulses with the period T=5 ms and filling factor of 10-25$\%$ to minimize heating effects at room temperature. The driving voltage was chosen so that the average excitation power density was in the order of hundreds of microwatts per $\mu m^2$(P$_{avg}=$100-700 $\mu$W/$\mu m^2$). The PL signal was collected through an apochromatic, high NA=0.65 microscope objective and measured with a standard Czerny-Turner spectrometer equipped with a CCD camera. Spatial filtering of the signal in the horizontal axis was performed by the entrance slit of the spectrometer, while the vertical axis filtration was done either by a multitrack feature of the CCD camera or, especially for the case of autocorrelation measurements, an additional slit placed in the intermediate imaging plane of a collimating setup added in the optical path. In Fig.\,\ref{fig:3} (c) we present examples of the obtained PL spectra of hBN narrow emission lines measured at room temperature. Each spectra corresponds to a different deposition process of the hBN powder onto the laser diode showcasing the possibility of stochastic tuning the emission energy between 600-800 nm, by consecutive stamping.  Typical linewidth values reside between 1.5 and 5 nm. To characterize the single-photon emission of the hBN narrow spectral lines we performed g$^{(2)}$ autocorrelation function measurements in a standard Hanbury Brown and Twiss configuration.

\begin{figure}[H]
    \centering
    \includegraphics{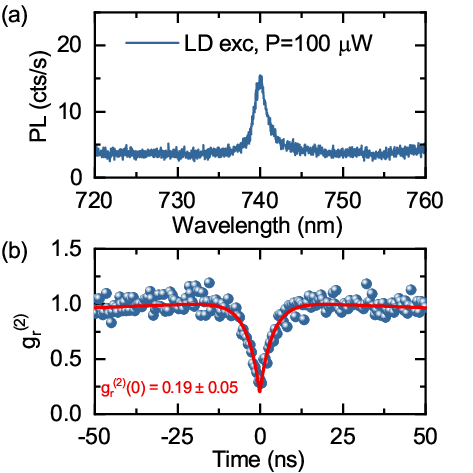}
    \caption{(a) Exemplary photoluminescence spectrum of the hBN-based emitter under LD excitation with an average optical power of 100 $\mu$W. (b) $g^{(2)}$ autocorrelation function of the photoluminescence signal of the emitter presented in (a). Fitted paramters: $A_1=1.32\pm0.03$, $A_2=0.631\pm0.005$, $t_1=3.9\pm0.2$ ns, $t_2=301\pm4$ ns}
    \label{fig:4}
\end{figure}

 In Fig.\,\ref{fig:4} we present the autocorrelation measurement (b) of an exemplary narrow emission line (a) obtained from the device. The $g^{(2)}$ function was fitted with a standard formula that takes into account a long timescale bunching effect often present in this material\cite{KoperskiOptComm18}.
 \begin{equation}
     g^{(2)}(\tau)=1-A_1 exp\left(-\frac{|\tau|}{t_1}\right)+A_2 exp\left(-\frac{|\tau|}{t_2}\right)
 \end{equation}
Here $A_1$ --- is the antibunching depth and $A_2$ --- bunching amplitude. $t_1$ corresponds to the emitter lifetime and $t_2$ is the timescale of the bunching effect.

For the case of the presented emitter, its lifetime is two orders of magnitude faster than the detected bunching timescale and nicely correlates with the literature-reported values of single nanoseconds\cite{KoperskiOptComm18}. After renormalization of $g^{(2)}_r$=$\frac{g^{(2)}}{1+A_2}$ we obtain the final value determining the single-photon character of the emission $g^{(2)}_r(0)=0.19\pm0.05$, which satisfies the general criteria for a single photon source ($g^{(2)}(0) < 0.5$). Having thus confirmed the room temperature single-photon emission from the device we now turn to the pulsed excitation schemes for on-demand generation of such emission.

\subsection{On-demand generation of single photons}\label{subsec23}
\begin{figure}[H]
    \centering
    \includegraphics{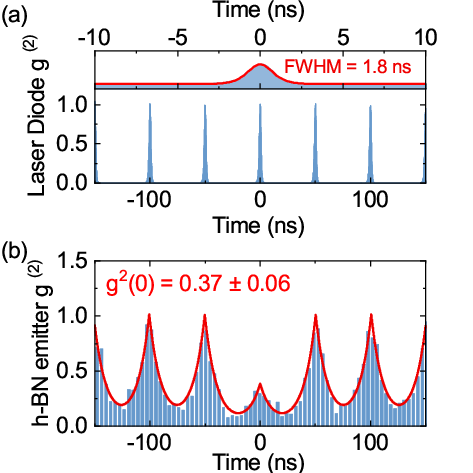}
    \caption{(a) g$^{(2)}$ autocorrelation function of the laser diode for pulsed driving with a period of 50 ns. The inset presents a close-up of a single autocorrelation peak of the FWHM of 1.8 ns.   (b) g$^{(2)}$ autocorrelation function of an exemplary hBN-based SPE excited by the laser diode in the pulsed operation mode. Exponential fit to the data yields the emitter lifetime of 11 ns.}
    \label{fig:5}
\end{figure}

To generate nanosecond electric impulses with sufficiently high currents we, once again, take advantage of the recent progress in the technology of nitride-based electronics \cite{MusumeciEnergies23} and use a home-made power supply in which we incorporate the commercially-available \textit{EPC8002} GaN field-effect transistor with a specialized driver \textit{LMG1020}. The presented laser diode is capable of pulsed operation where the temporal width of the impulses does not exceed the typical lifetime of studied emitters. This property, crucial for the on-demand generation of single photons, was verified with autocorrelation measurements of the laser emission. In Fig.\,\ref{fig:5}(a) we show the $g^{(2)}$ function of the LD under pulsed driving with a period T=50 ns. The resultant autocorrelation peaks display a Full-Width-Half-Maximum of 1.8 ns and decay slope of $t_{dec}=0.56$ ns. The operating speed of the LD is limited by its capacitance, which can be estimated to be $C\approx10$ pF. Here, we assumed an RC circuit and considered the impedance of the utilized coaxial cables of $R=50$ Ohm.

In Fig.\,\ref{fig:5}(b) we show an example of the autocorrelation function of an hBN emitter's photoluminescence excited by the LD under pulsed driving. Exponential peaks fitted to the data yield a lifetime of 11 ns and $g^{(2)}(0)=0.37\pm0.06$. Given that this value satisfies the criterion for single-photon emission we thus confirm the generation of a single photon per excitation pulse from the LD device.

\section{Conclusions}\label{sec3}
With the ever-increasing demand for single-photon-generating devices driven by the rapid development of photon-based quantum technologies, layered semiconductors have emerged as promising materials and have begun to take center stage in nanophotonics research. In this work, we overcome one of the most limiting factors for producing hBN-based optical devices by manufacturing structures that utilize a novel bottom-to-top excitation scheme. We presented that the produced hybrid device, even at this early stage of research, when compared to the historical progress in QD-based technologies, already provides on-demand generation of single photons at room temperature and without the necessity for external laser excitation. The achieved values of the zero-delay autocorrelation function indicate, however, the requirement for further optimization to match the industry standards. This may be partly due to the selected material of hBN nanocrystals. Its manual deposition on the laser facet provides luminescence from different SPEs in the 600-800 nm range and, while being time/cost-efficient, it does not employ any further optimization of the emitter's position or annealing treatments\cite{BreitweiserACSPhot20}. In the future, certain methods for deterministic generation of efficient single-photon emitting sites could be integrated to improve both the efficiency and scalability of the production of these devices\cite{CianciNanoFut24}. This is particularly promising while considering the recently reported achievement of visible Hong-Ou-Mandel effects in the two-photon interference of hBN SPE's signal, which is the next big step towards the generation of indistinguishable photons\cite{FournierPhysRevApplied23}.

\medskip
\textbf{Supporting Information} \par 
Supporting Information is available from the Wiley Online Library or from the author.

\medskip

\medskip
\textbf{Funding}\par
This work was supported by the National Science Centre, Poland under projects: 2020/39/B/ST3/03251, 2022/45/B/ST7/03964, 2020/39/B/ST7/02945, and Foundation for Polish Science under TEAM-TECH POIR.04.04.00-00-210C/16-00 and TEAM POIR.04.04.00-00-1A18/16-01 (ATOMOPTO) projects.

\end{document}